\documentclass[%
onecolumn,
superscriptaddress,
amsmath,amssymb,
aps,
]{revtex4-1}

\usepackage{graphicx}
\usepackage{dcolumn}
\usepackage{bm}

\begin{document}

\preprint{APS/123-QED}

\title{Unsupervised organization of cervical cells using high resolution digital holographic microscopy}

\author{Jyoti Mangal} \thanks{Jyoti Mangal and Rashi Monga have contributed equally to this work.}
\affiliation{Department of Physics, Indian Institute of Technology Delhi, New Delhi 110016 India}
\author{Rashi Monga} \thanks{Jyoti Mangal and Rashi Monga have contributed equally to this work.}
\affiliation{Department of Physics, Indian Institute of Technology Delhi, New Delhi 110016 India}
\author{Sandeep R. Mathur}
\affiliation{Department of Pathology, All India Institute of Medical Sciences, New Delhi 110039 India}
\author{Amit K. Dinda}
\affiliation{Department of Pathology, All India Institute of Medical Sciences, New Delhi 110039 India}
\author{Joby Joseph}
\affiliation{Department of Physics, Indian Institute of Technology Delhi, New Delhi 110016 India}
\author{Sarita Ahlawat}
\affiliation{Phase Laboratories Pvt. Ltd., TBIU-IIT Delhi, New Delhi 110016 India}
\author{Kedar Khare}
\affiliation{Department of Physics, Indian Institute of Technology Delhi, New Delhi 110016 India}
\email{kedark@physics.iitd.ac.in}

\begin{abstract}
We report results on unsupervised organization of cervical cells using microscopy of Pap-smear samples in brightfield (3-channel colour) as well as high resolution quantitative phase imaging modalities. A number of morphological parameters are measured for each of the $1450$ cell nuclei (from $10$ woman subjects) imaged in this study. The Principal Component Analysis (PCA) methodology applied to this data shows that the cell image clustering performance improves significantly when brightfield as well as phase information is utilized for PCA as compared to when brightfield-only information is used. The results point to the feasibility of an image-based tool that will be able to mark suspicious cells for further examination by the pathologist. More importantly our results suggest that the information in quantitative phase images of cells that is typically not used in clinical practice is valuable for automated cell classification applications in general. 
\end{abstract}

\maketitle

\section{Introduction}\label{sec1}
Visual examination of cell samples using a microscope is a commonly used diagnostic methodology world over. The pathologists typically rely on their subjective interpretation of various cell features or other diagnostic markers that they learn over years of practice. At a standardized resolution using a $40$x microscope objective, imaging of a typical whole slide amounts to tens of gigabytes of data. Visual inspection of this image data is time consuming and also suffers from inter-observer variability. With the advent of digital imaging technologies like whole slide scanner, a lot of effort is underway to combine digital imaging with automated machine learning methods for cell classification that can help improve the accuracy of diagnostic tasks. The success of automated diagnostic tools critically depends on the amount of non-redundant information that can be made available to the classification algorithms. In the present work we analyze cervical cell image data obtained using a microscope operating in both bright field as well as quantitative phase imaging modes. 

Cervical cancer is the second most common cancer among women worldwide \cite{nccc}. Although mortality rate has been known to decrease with regular screening \cite{Andrae2012}, the situation is still worse in developing countries where there is general lack of resources and skilled clinicians \cite{Cronje2014}. One of the peculiar aspects of cervical cancer is that it may take as long as a decade to develop fully \cite{Kiviat1996}, as a result, early intervention with an effective screening procedure can boost the survival rate considerably. 
Papanicolaou test or Pap test is one of the popular screening tests in which cells from the cervix are collected using a brush/spatula and observed under a microscope for the detection of pre-cancerous or cancerous cells \cite{Pap}. Microscopic examination of biopsy sample is considered as the ultimate confirmatory test. The pathologists scan the microscope slides to study the cytological changes in the nuclear profile of the cells as the cells transition from normal to cancerous. Based on the nuclear profile, the abnormal patient sample is categorized into one of the following main stages- Atypical Squamous Cells of Undetermined Significance (ASC-US), Low-grade Squamous Intraepithelial Lesions (LSIL), High-grade Squamous Intraepithelial Lesions (HSIL) and Adenocarcinoma \cite{MathurBook}. Due to varying methods of Pap slide sample preparation and a number of qualitative characteristics of the cell nuclei that are used by pathologists in visually inspecting the cell samples, the cervical cell classification is subjective in nature and as a result the typical early detection rate is known to be low \cite{DetectRate}. In order to address these difficulties, several prior efforts have been made to combine images from the Pap slides with machine learning ideas to develop automated classification strategies \cite{Marinakis, Song, Chankong, Bora, DeepPap, RandomForest}. Annotated benchmark databases like the Herlev database \cite{Herlev} consisting of $917$ brightfield images of cervical cells belonging to various categories are also available. These studies fully rely on features obtained from brightfield 2D images of cervical cells. More recently, newer benchmark databases have also been made available for automated cell classification studies \cite{sipakmed}. These databases and the related classification studies fully rely on features obtained from brightfield 2D images of cervical cells. Further majority of them require supervised learning where individual cells in a training dataset are labeled by a pathologist in order to train classification algorithms. Our aim in this paper is to use additional non-redundant information in the form of quantitative phase of the cervical cell nuclei as obtained using a high resolution digital holographic microscope for cell classification. A preliminary study of Digital Holographic Microscope (DHM) based imaging of cervical cells has been reported \cite{nazim} showing the importance of this methodology for cervical cell screening. Quantitative assessment of cancer cell morphology using a DHM has also been reported \cite{raub}. The present work has several novel aspects including phase image reconstruction methodology as well as analysis using a number of phase based features as we will discuss in this paper.

For the present study, we recorded both bright field images and image plane holograms of cervical cells from the Pap-smear slides using a digital holographic microscopy (DHM) operating in dual modes (brightfield as well as quantitative phase). The brightfield images were obtained using a white LED illumination while the phase images were recovered from digital holograms recorded using a low-power diode laser illumination. DHM is a mature technology \cite{Kim, Schnars, Kemper} with two standard phase reconstruction methodologies employed in the literature so far. The first methodology involves single shot off-axis digital holography. The off-axis holograms are typically analyzed using the Fourier Transform method \cite{Takeda}. This method inherently involves a low-pass filtering operation thus giving rise to phase images with lower resolution as compared to sensor array resolution capability. The second methodology involves phase shifting \cite{yamaguchi} which is a multi-frame approach requiring stringent vibration isolation requirements that may be difficult to achieve in a practical clinical environment. For phase reconstruction from DHM data we use an optimization method that is capable of providing single-shot full diffraction-limited resolution over a selected region-of-interest (ROI). Over recent years, our work has demonstrated the superiority of this optimization method over the traditional single shot Fourier transform method as well as the phase shifting method \cite{{Khare2013}, {Khare2014}, {PRA2015}, {josaa2017}, {Singh2018}}. As a result, a region-of-interest enclosing the cell nucleus can be reconstructed with full diffraction limited resolution in single-shot off-axis holographic configuration.   

Based on the description of qualitative parameters commonly used by pathologists \cite{MathurBook}, a number of quantitative parameters were designed and their numerical values for the cell nuclei under study were evaluated using both the brightfield as well as the quantitative phase images. The numerical feature data so created was then analyzed using Principal Component Analysis (PCA) to organize the cell data in an unsupervised manner. The clustering and organization of the cell image data is valuable for clinicians in order to efficiently utilize the limited time available to them for examining the entire slide. An important finding of our study is that PCA using both brightfield as well as phase information is sigificantly superior at cell data organization as compared to the PCA performed with brightfield information alone. Quantitative phase information is thus seen to provide valuable additional information for the purpose of cell data organization. The results are important in the context of building an automated cell classification tool for assisting the pathologists.

The paper is organized as follows. In section 2, we briefly discuss the background information on extracting high resolution phase information from single-shot image plane holograms of the Pap-smear slides. In section 3, we describe the methodology for feature extraction from the cell nuclei. In section 4, we show results of our analysis using PCA and finally in section 5, we provide our concluding remarks as well as future directions.

\section{High Resolution Digital Holographic Microscopy} 
When a spatially coherent beam of light passes through a nearly transparent cell, the amplitude or intensity of light is nearly unchanged but the phase of the transmitted wave front gets modified due to the finite cell thickness and local refractive index variations. The phase $\phi (x,y)$ of the transmitted monochromatic wave (wavelength $\lambda$) through a cell may be described as:
\begin{equation}\label{eq:opd}
\phi (x,y) = \frac{2 \pi}{\lambda} \int dz \; n_r(x, y, z)\, ,
\end{equation}  
where $n_r(x,y,z)$ is the real part of the refractive index profile of the cell. 

The resulting wave front carries an imprint of the cell$'$s optical thickness which can code new information on cell structure that is not available with the typical brightfield or fluorescence imaging modalities. The refractive index of the cell can be represented as $n(x,y,z) = n_r (x,y,z) + i n_i (x,y,z)$ where the real part {$n_r$} of the refractive index forms the phase image and the imaginary part $n_i$, which is responsible for light absorption is predominantly responsible for contrast in the bright-field image. In our experiments we recorded bright field images of the cell samples using a white LED illumination followed by an image plane hologram using a low power laser illumination. Since the phase information in the resulting image is not derivable from the bright field image, we ultimately obtain four independent channels of information viz. red, green and blue channels from brightfield image and phase image $\phi(x,y)$ as the fourth channel. From Eq. (\ref{eq:opd}) we observe that the phase information can be related to the optical height of the cell or its depth dimension. The independence between phase and the amplitude means that the phase image adds new non-redundant information to the cell image data which is important for effective classification.

\begin{figure}[tb]
\centering
\includegraphics[width = 0.9\textwidth]{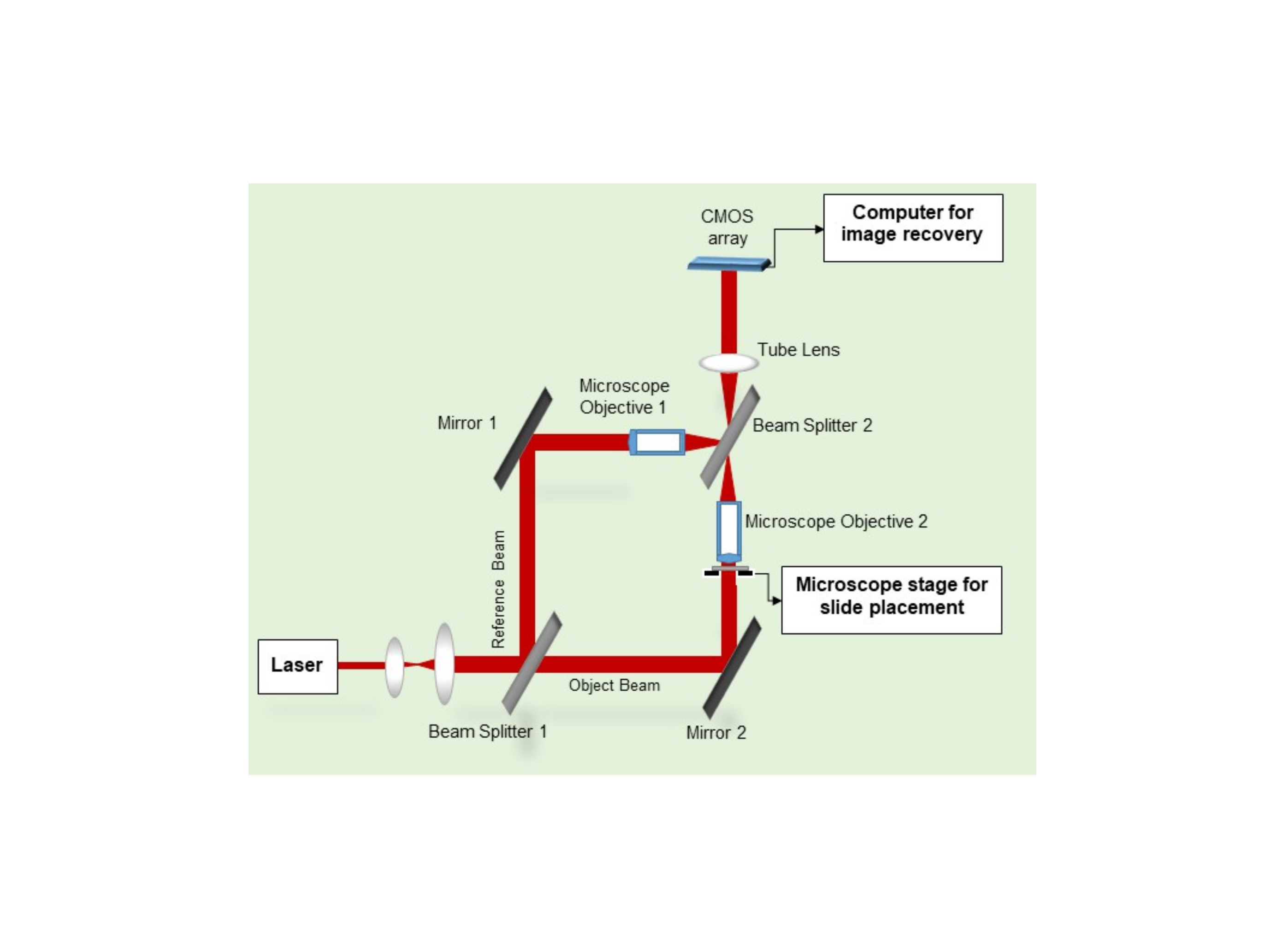}
\caption{Digital Holographic Microscope (DHM) setup depicting the Mach-Zehnder configuration used with balanced path length in the reference and object arms.}
\label{MZI}
\end{figure}

Digital holographic microscopy is a technique to obtain the phase information that involves recording of a hologram or interference pattern between a reference beam and object beam- both obtained from the same source on a sensor array (CCD or CMOS camera). The phase information is extracted by reconstructing the hologram using numerical computational methods \cite{{Kim}, {Schnars}}. We used a Digital Holographic Microscope (DHM) set up (Model: HO-DHM-UT-01, Make: Holmarc Opto-Mechatronics Pvt. Ltd., Kochi, India) with balanced 40x infinity corrected objectives to obtain both bright field and corresponding hologram. A CMOS camera (Model: uEye 3070CP, $2056 \times 1542$ pixels of size 3.5$\mu$m; Make: IDS-imaging, Germany) was used to record the images. The DHM system used a Mach-Zehnder configuration for obtaining the holograms as illustrated in Fig. \ref{MZI}. A low power collimated laser beam ($3$ mW, $\lambda$ = $650$ nm) is first split into two parts by a beam-splitter. The reference and object arms of the interferometer are balanced. A small tilt in the beam-splitter 2 makes it possible to get a tilted plane reference beam $R(x,y)$ at the sensor plane, whereas the infinity imaging system provides a magnified image field $O(x,y)$ (corresponding to the cell sample plane) at the sensor plane. A white light LED illumination that can be used by blocking the laser beam was separately provided in the DHM system (not shown in Fig. \ref{MZI}), so that the sample slide plane can be imaged onto the sensor in the brightfield mode.  Due to the temporal coherence of the laser, the resultant hologram intensity pattern $H(x,y)$ recorded on the sensor can be described as: 
\begin{equation}
H(x,y) = | R(x,y) + O(x,y) |^2 ,
\end{equation}
The reference beam $R(x,y)$ is assumed to be of the form:
\begin{equation}
R(x, y) = R_o e^ {i2\pi (f_{ox} x + f_ {oy} y)}
\end{equation}
where $f_{ox}$ and $f_ {oy}$ are the spatial frequencies in the definition of the plane reference beam in $x$ and $y$ directions respectively. The reference beam can be determined as a calibration step using the straight line fringe pattern obtained when there is no sample placed in the object arm of the interferometer.
A popular method for analyzing single-shot off axis holograms is the Fourier transform method \cite{Takeda} which performs a spatial filtering operation of the hologram to extract information near the cross-term peak in Fourier space of the hologram. Although this method single-shot, it is inherently a low-pass filtering operation. The image plane holograms that we record for our studies have sharp edge-like features that are blurred when the Fourier filtering operation is used for processing the off-axis holograms. As shown in our recent works, the optimization method can recover phase images with the same theoretical diffraction-limited resolution as the bright-field images\cite{Khare2013, josaa2017, Singh2018} as well as superior noise performance \cite{PRA2015} for given hardware configuration. In the optimization method, the recovery of the image field $O(x, y)$ from a single-shot hologram $H(x, y)$ is modeled as a problem of minimization of a cost function of the form:
\begin{equation}
C(O, O^*) = C_1(O, O^*) + C_2(O, O^*) 
\end{equation}
where 
\begin{align}
C_1(O, O^*) &= \|{H - ( |R|^2 + |O|^2 + R^*O + RO^* )}\|^2 ,
\end{align}
and
\begin{align}
C_2(O, O^*) &= \sum_{j = all pixels} \bigg[\sqrt{1 + \frac{ |\nabla O_j|^2}{\delta ^2}} - 1 \bigg].
\end{align}
The first term above is an L2-norm squared data fitting term and the second term is the modified Huber penalty.
The parameter $\delta$ in the penalty term is made proportional to the median of gradient magnitudes $|\nabla O_j|$. The penalty function thus behaves like an edge preserving total variation (TV) penalty for pixels with $|\nabla O_j| >> \delta$ and as a smoothing penalty for $|\nabla O_j| << \delta$. The modified Huber penalty allows us to retain sharp edge like features as well as smooth greyscale features in the reconstructed phase image. In practice we use an alternating minimization framework \cite{Singh2018, josaa2017} where the solution is obtained by adaptively minimizing the two terms of the cost function. Furthermore, this technique 
operates directly in the image space so that the iterative procedure may be carried over only a subset of all the recorded hologram pixels thereby enabling region-of-interest (ROI) image reconstruction \cite{Singh2018} which is less computationally intensive leading to near real-time phase reconstruction of the selected ROI. The ROI in our work is typically selected centered around a cell nucleus and the corresponding information from both the brightfield and phase images is used for extracting morphological features of the cell nuclei.
\begin{figure}[tb]
\centering
\includegraphics[width = \textwidth]{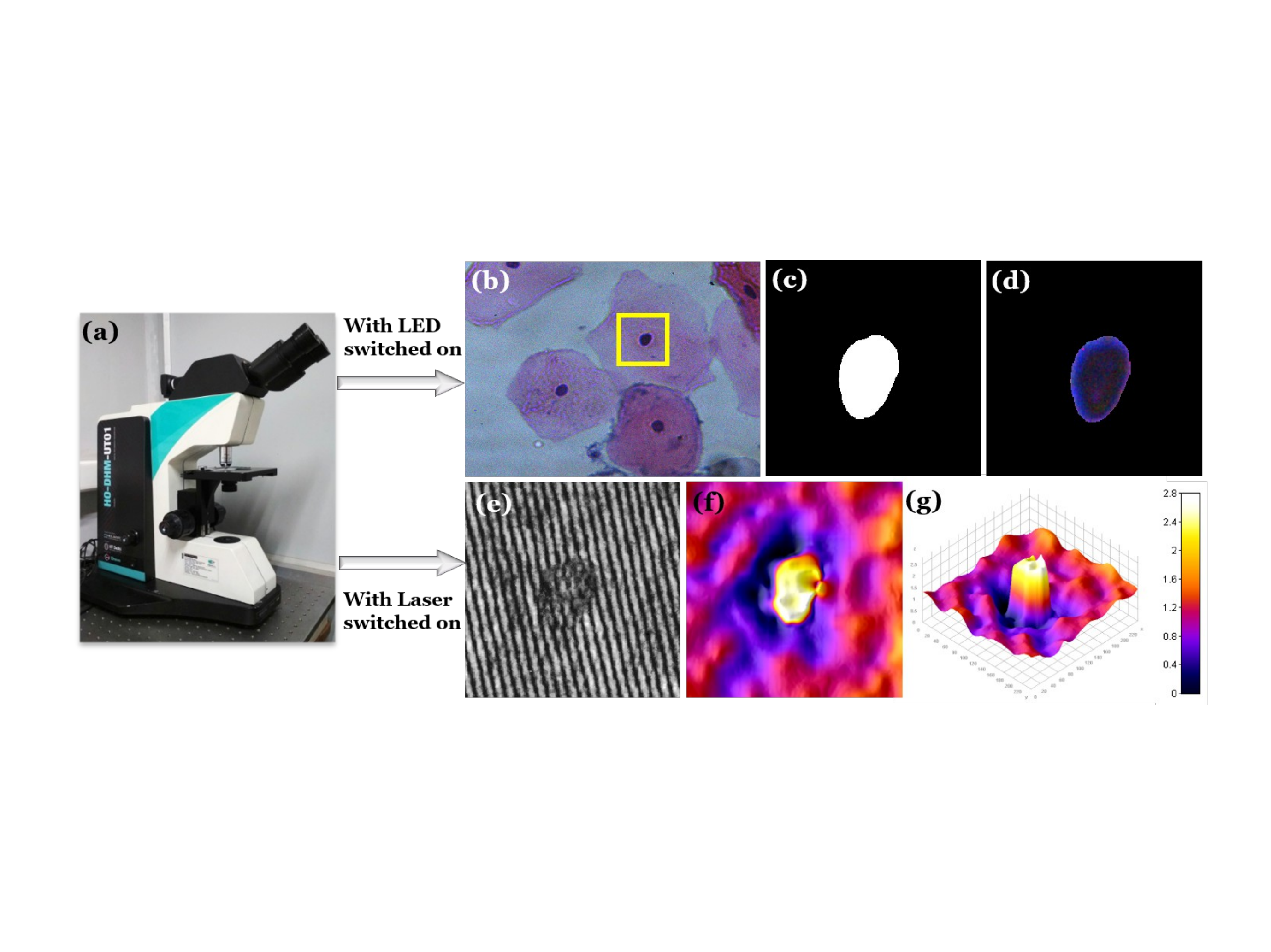}
\caption{Flowchart depicting the series of processing steps in order to obtain the cervical cell parameters: (a) Digital Holographic Microscope, (b) Bright field image with ROI indicated by box, (c) Binary Mask Image ROI, (d) Mask matrix multiplied with bright field ROI, (e) Hologram ROI, (f) Phase image, and (g) 3-D phase image rendering of the nucleus}
\label{Chain}
\end{figure}
Having reconstructed the object field $O(x,y)$, the phase $\phi(x,y)$ in the reconstructed holographic images is usually expressed as:
\begin{equation}
\phi(x,y) = \arctan\big( \frac{Im[O(x,y)]}{Re[O(x,y)]} \big),
\end{equation}
where $Re[O(x,y)]$ and $Im[O(x,y)]$ denote the real and imaginary parts of the object field recovered from the hologram. The phase function is thus wrapped to the interval $[-\pi, \pi]$ \cite{Malacara}. The $-\pi$ to $\pi$ jumps in the resultant phase map are however artificial and to determine accurate phase profile of cell nuclei, the reconstructed image needs to be unwrapped. For unwrapping the phase we use a transport-of-intensity based technique \cite{Pandey} introduced by us recently. The unwrapped phase map can be used in the definition in Eq. (\ref{eq:opd}) to relate it to the optical depth associated with the cell nucleus. In our study the nuclei of interest were selected to obtain an ROI of $256 \times 256$ pixels from a $2056 \times 1542$ pixels hologram frame. The reconstruction and unwrapping of phase profiles were implemented in MATLAB. 
\section{Cell Nuclei Measurement}
\begin{figure}[tb]
\centering
\includegraphics[width = 0.5\textwidth]{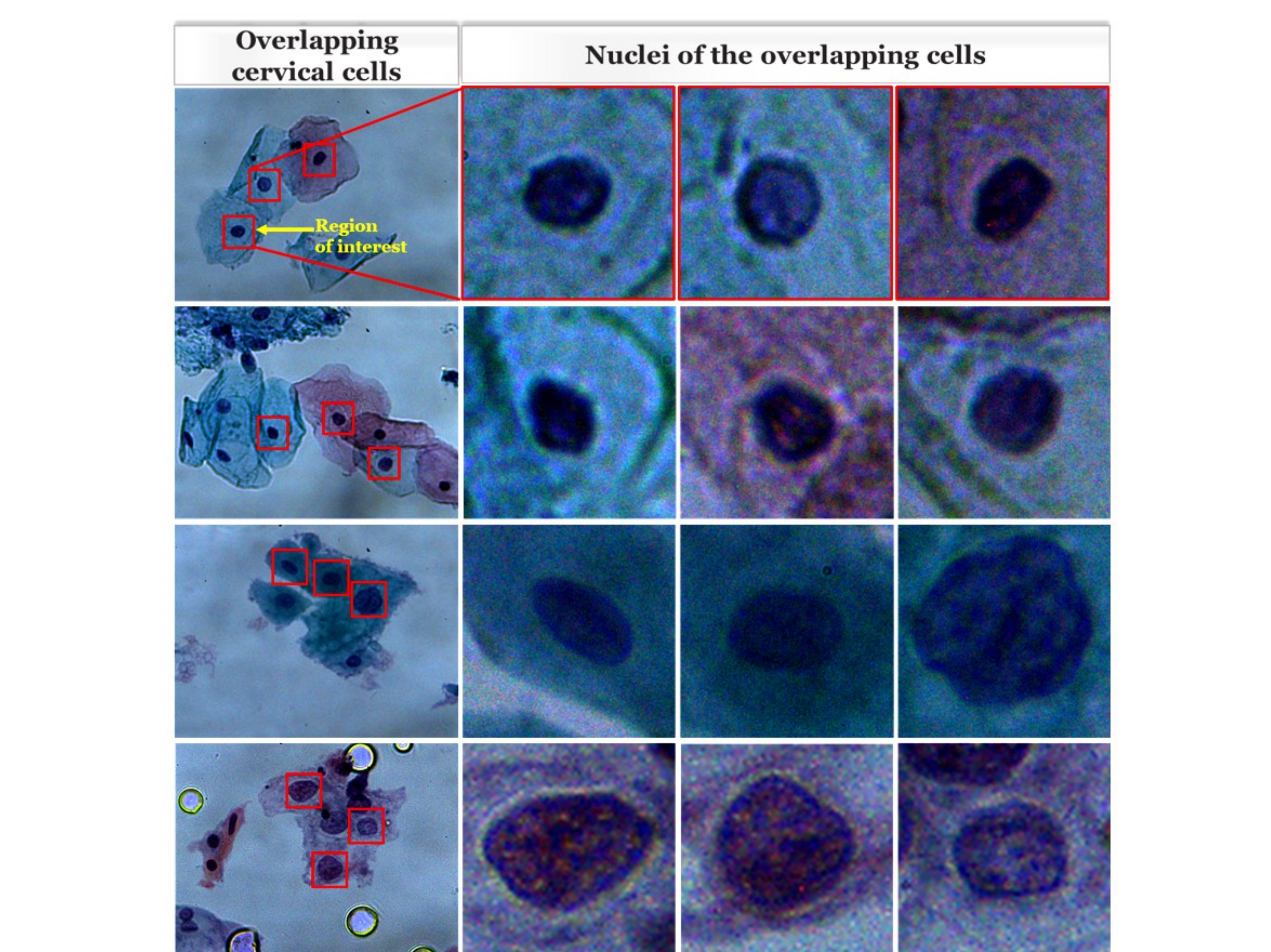}
\caption{Brightfield images from our dataset depicting cervical cells with overlapping cytoplasm. The nuclei of the cells can however be seen to be non-overlapping. Overlapping cells constituted $60 - 70 \%$ of cell population.}
\label{NCR}
\end{figure}

The imaging of cervical cells was performed using a DHM system as shown in Fig. 2(a). For a fixed position of sample we recorded the brightfield as well as interferometric data corresponding to the cells under study. A GUI interface was designed such that the user can select the location of individual cell nucleus with a mouse-click in the brightfield image as shown in Fig. \ref{Chain}(b). A $256 \times 256$ ROI centered on the nucleus was first selected from the brightfield image. In order to define the nucleus boundary, the following pre-processing steps were performed on this ROI. This ROI of the brightfield image was first transformed from RGB to the YUV color space and the Y-channel was selected to form a grey-level image. This was done to enhance the brightness and contrast of the image so as to get the best visual representation. Median filter was applied on this image data for efficient noise removal. Cell segmentation was finally carried out on the preprocessed ROI image using K-Means clustering. We partitioned the ROI matrix into two clusters according to their pixel intensity values after pre-processing. For the desired cell segmentation, the nucleus region pixels were assigned the value of $1$, and the surrounding region is assigned the value of $0$, thereby forming a binary mask matrix as illustrated in Fig. \ref{Chain} (c). This binary mask matrix is placed on the brightfield and phase ROI images to select the desired pixels in the ROI corresponding to the cell nucleus. The masked brightfield image is shown in Fig. \ref{Chain} (d). The corresponding ROI of the digital hologram is shown in Fig. \ref{Chain}(e) which shows off-axis straight line interference fringes which are modulated at the location of the cell nucleus. The single-shot optimization methodology described in the section $2$ was applied to this ROI hologram and the resultant phase image of the nucleus is shown in Fig. \ref{Chain}(f). Since the phase image can be associated with the optical thickness of the cell, a 3D rendering of the phase map is also shown in Fig. \ref{Chain}(g) for developing a visual understanding of the phase image. The mask in Fig. \ref{Chain}(c) is further used to select the cell nucleus in the phase image as well. The mask regions in the brightfield as well as phase images allow us to effectively compute various morphological parameters associated with the cell nucleus.
It is well known that the pathologists examine the brightfield images of cervical cells under a microscope
and use several qualitative parameters associated with the cell nuclei such as abnormal chromatin pattern, disproportionate nuclear enlargement, irregular nuclear membrane, nuclear hyperchromasia, reduced cytoplasm, and multiple abnormal nuclei. Based on the combination and intensity of these markers, the patient sample is categorized into normal or various stages of pre-cancerous cells \cite{MathurBook}. Given a set of cell images,
this visual inspection methodology is subjective and it is very important to organize the cell image data so that cells of similar type can be grouped. Such organization of cell image data can significantly ease the pathologists' workload and they can pay more attention to the abnormal cases. The full-resolution phase images of cell nuclei available with our methodology represent a new physical information that the pathologists cannot access with a brightfield microscope. Since phase information effectively translates to cell morphology, phase captures valuable information regarding the distribution of material inside the nucleus. Our aim in this work is therefore to combine information from brightfield as well as phase images for effective organization of cell image data. We proceeded to compute a number of quantitative parameters from the selected masked regions representing the cell nuclei. The numerical parameters used by us are listed in Table \ref{tab1} along with brief description. In this table, \emph{M} denotes the binary mask matrix and \emph{$\phi$} denotes the unwrapped phase image ROI matrix. The parameters are designed after a study of the various qualitative aspects of the cell images used by pathologists \cite{MathurBook}. The numerical values of these parameters can be calculated in a straightforward manner from the ROI images. We emphasize once again that the parameters measured from phase images cannot be derived from usual brightfield images and therefore represent new information.
\begin{center}
\begin{table*}[t]
\caption{Parameters evaluated for each nucleus.}\label{tab1}
\centering
\begin{tabular*}{500pt}{@{\extracolsep \fill}p{155pt}p{335pt}@{\extracolsep \fill}}
\toprule
\bfseries Bright Field Parameters & \bfseries Description \\
\hline
Area &  $= \sum_{i,j} M_{i,j}$ (Number of pixels in the masked region) \\ [0.5ex]

Perimeter &	Compute the gradient of the mask, G = $\sqrt{(\frac{\partial M}{\partial x})^2 +(\frac{\partial M}{\partial y})^2}$
Then, Perimeter = $\sum_{i,j} G_{i,j}$ \\ [0.5ex]

Perimeter per unit area &	Perimeter per unit area$;$ Accounts for the irregularity in the boundary of the nucleus. \\ [0.5ex]

Mean intensity &	Mean of intensity values\\ [0.5ex]

Mean intensity in red, green and blue	& Mean of intensity values of red, green and blue channels computed individually \\ [0.5ex]

Variance of intensity in red, green and blue &	Variance of intensity values computed individually for red, green and blue channels each \\ [0.5ex]

\toprule
\bfseries Phase Parameters & \bfseries Description \\
\hline
Optical Volume &	Sum of all the pixel values of the masked phase image of the nucleus $ = \sum_{i,j} \phi_{i,j} M_{i,j}$ \\ [0.5ex]

Roughness/Area &	Compute the gradient of the phase image, $ P = \sqrt{(\frac{\partial \phi}{\partial x})^2 +(\frac{\partial \phi}{\partial y})^2}$ Then, Roughness $ = \sum_{i,j} \phi_{i,j} M_{i,j} / Area$ \\      [0.5ex]

Roughness/Area at different scales &	Both phase image and mask are scaled down by factors of $\frac{3}{4}$, $\frac{1}{2}$, $\frac{1}{4}$, $\frac{1}{8}$ using bicubic interpolation and roughness is calculated at all these scales individually \\ [0.5ex]

Variance	& Variance of the phase values corresponding to the nucleus \\ [0.5ex]

Shift in the centroid  & Euclidean distance between the phase image centroid and geometric centroid based on mask M is computed. This value describes if the material in the nucleus is distributed uniformly.\\ [0.5ex]

Moment of inertia &	Moment of Inertia of the masked phase image is evaluated to understand the distribution of nuclear material around its center of mass given by $ \sum_{i,j} \phi_{i,j} r^2 +  \sum_{i,j} \phi_{i,j} d^2$, where r denotes the distance between the origin to the pixel (i, j) and d indicates the distance between the origin and phase centroid. Here, the first term indicates the moment of inertia about the normal axis through the origin. The last term is included as per parallel axis theorem to evaluate the moment of inertia about the phase centroid.\\  [0.5ex]
\hline
\end{tabular*}
\end{table*}
\end{center}
One of the common parameters that pathologists look for while examining cells is the ratio of the size of the nucleus in a given cell to the size of the cytoplasm. The nucleus-to-cytoplasmic ratio (NCR) is typically known to increase as a cell transitions from healthy to malignant \cite{MathurBook}. NCR is typically assessed visually by pathologists and a larger NCR value may be associated with abnormal chromatin activity often seen in pre-cancerous and cancerous cervical cells.  In addition to the subjectivity associated with the visual inspection of NCR, a large fraction ($ ~ 60 - 70 \% $) of the cells in our Pap smear samples were observed to have cell clusters with overlapping cytoplasm. Figure \ref{NCR} shows a few typical brightfield images of overlapping cervical cells used in our study. While NCR may still be approximately assessed visually, automated segmentation of individual cells with overlapping areas is difficult with the use of standard image segmentation tools. In \cite{Genctav} the total cytoplasm area of a cluster has been divided by number of nuclei in a cluster to estimate the average cytoplasm area for each cell, which is not necessarily an accurate measure for calculating NCR. The visual assessment of NCR has been shown to have a high inter-observer variability \cite{Schmidt} thus making it a somewhat less robust criterion for cell classification. As can be seen in Fig. \ref{NCR} the nuclei of majority of the cells usually did not overlap and are easily separated by intensity based segmentation. We have therefore omitted the NCR parameter for this study and used the parameters as listed in Table \ref{tab1} for PCA as described in the next section. 
\section{PCA analysis of cell data}
\begin{figure}[tb]
\centering
\includegraphics[width = 0.9\textwidth]{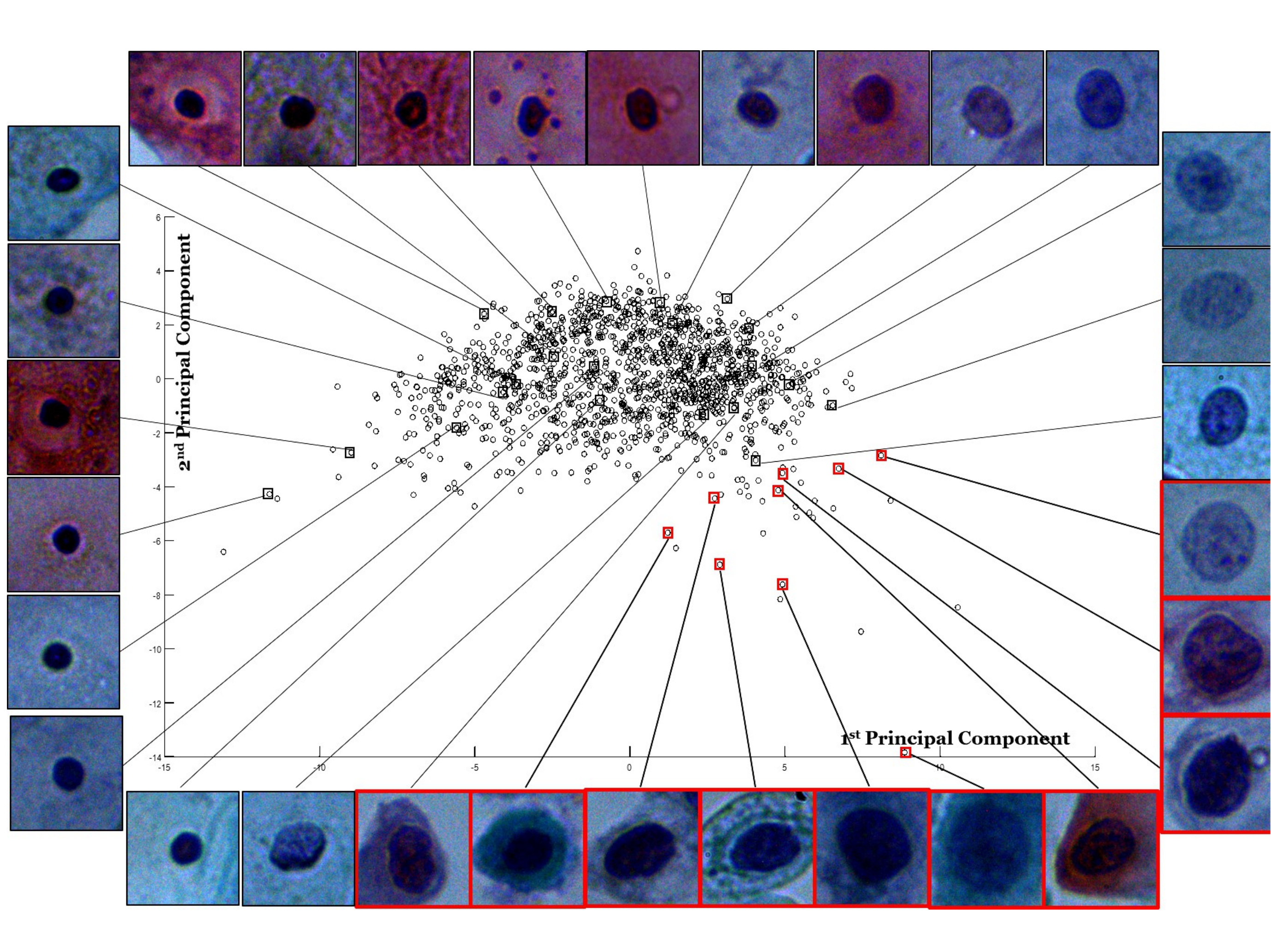}
\caption{Representation of cell data using first two PCA vectors obtained using data corresponding to $21$ measurements obtained from brightfield as well as phase images of $1450$ cell nuclei. The brightfield images of a representative set of $30$ cell nuclei are also shown on the plot. The representation shows natural organization of cell data.}
\label{fig:allpca}
\end{figure}

A data set of cervical cell images in brightfield and phase modalities consisting of $1450$ cell nuclei was recorded using our DHM system and $21$ parameters listed in table\ref{tab1} were evaluated for each cell nucleus. The number of cells used here is almost $1.5$ times the number of cell images in the standard Herlev database \cite{Herlev}. The resultant data was analyzed using principal component analysis (PCA) \cite{PCA} for performing its unsupervised organization. PCA is mathematically defined as an orthogonal linear transformation where a data set described by correlated coordinate axes is transformed into a new coordinate system spanned by the orthogonal axes called principal components. This methodology does not require prior labeling of individual cells and is thus completely unbiased. The transformation is defined in such a way that the first principal component captures the largest possible variance in data, and each succeeding component in turn captures the variance in the data in reducing order of importance under the constraint that it is orthogonal to the preceding components. For the problem at hand the data corresponding to the $21$ measurements for the $1450$ nuclei can be arranged as a matrix $\textbf{D}$ of size $1450 \times 21$ where the columns represent mean adjusted and normalized (by standard deviation) values of the parameters of interest. The principal vectors are obtained by solving the eigenvalue problem:
\begin{equation}
\textbf{D}^{T} \textbf{D} \textbf{u}_k = \mu_k \textbf{u}_k;
\end{equation}
corresponding to the correlation matrix of $\textbf{D}^{T} \textbf{D}$ associated with the normalized and mean adjusted data. Here the superscript $T$ stands for the transpose of the matrix $\bf{D}$. The eigenvectors $\textbf{u}_k$ are mutually orthogonal and form the principal directions for representation of the data. The eigenvalues $\mu_k$represent the variance along the corresponding eigenvector $\textbf{u}_k$. Only the first few eigenvectors are typically significant and hence the original $21$-dimensional data space can be reduced effectively to the space spanned by the first few principal vectors. Denoting the eigenvector matrix whose columns denote the PCA eigenvectors $\textbf{u}_k$ for data corresponding to $n$ cell nuclei by ${\bf{U}^{(n)}}$, we look for the incremental change in the eigenvector matrix as $(n+1)^{th}$ data point is added to the PCA eigenvector calculation. After randomizing the cell sequence in the tabulated measurements, we found that the relative change in the eigenvector matrix as measured by  
\begin{equation}
\frac{||\textbf{U}^{(n+1)} - \textbf{U}^{(n)}||}{||\textbf{U}^{(n)} ||} \leq 0.005,
\end{equation} 
is insignificant for $n$ larger than $400$. Here the notation $|| ... ||$ represents the Frobenius norm of the matrix quantity inside. The PCA vectors may therefore be considered to be well stabilized. The Kaiser-Meyer-Olkin (KMO) score \cite{KMO, KMO1} for the data corresponding to the $1450$ nuclei was found to be $0.80$ implying very good sampling adequacy of the data for performing PCA. The first two principal components of the data set were found to explain $71 \%$  of the variation in the data. A scatter plot of the cell data represented using the first two PCA vectors is shown in Fig. \ref{fig:allpca}. Each of the points in this scatter plot represents a cell nucleus. For ease of interpreting the cells that are represented on the plot, the brightfield images of a selected set of $30$ representative nuclei are shown on the plot. We observe that the cells of similar kind have automatically been organized close to each other. The cell types show a natural gradient across the first principal component of the plot as identified by a practicing pathologist (S.M.). The extreme left of the cluster is dominated by the class of superficial cells with pyknotic nuclei, whereas the extreme right is dominated by vesicular nuclei of intermediate cells. In particular the cells on the bottom right corner (indicated by red boundary) were identified as abnormal requiring further attention for diagnostic purposes. We emphasize here that the organization of the cell data on the PCA plot required no prior input from the pathologist and was performed in a completely blinded and unsupervised fashion.

\begin{figure}[tb]
\centering
\includegraphics[width = 0.9\textwidth]{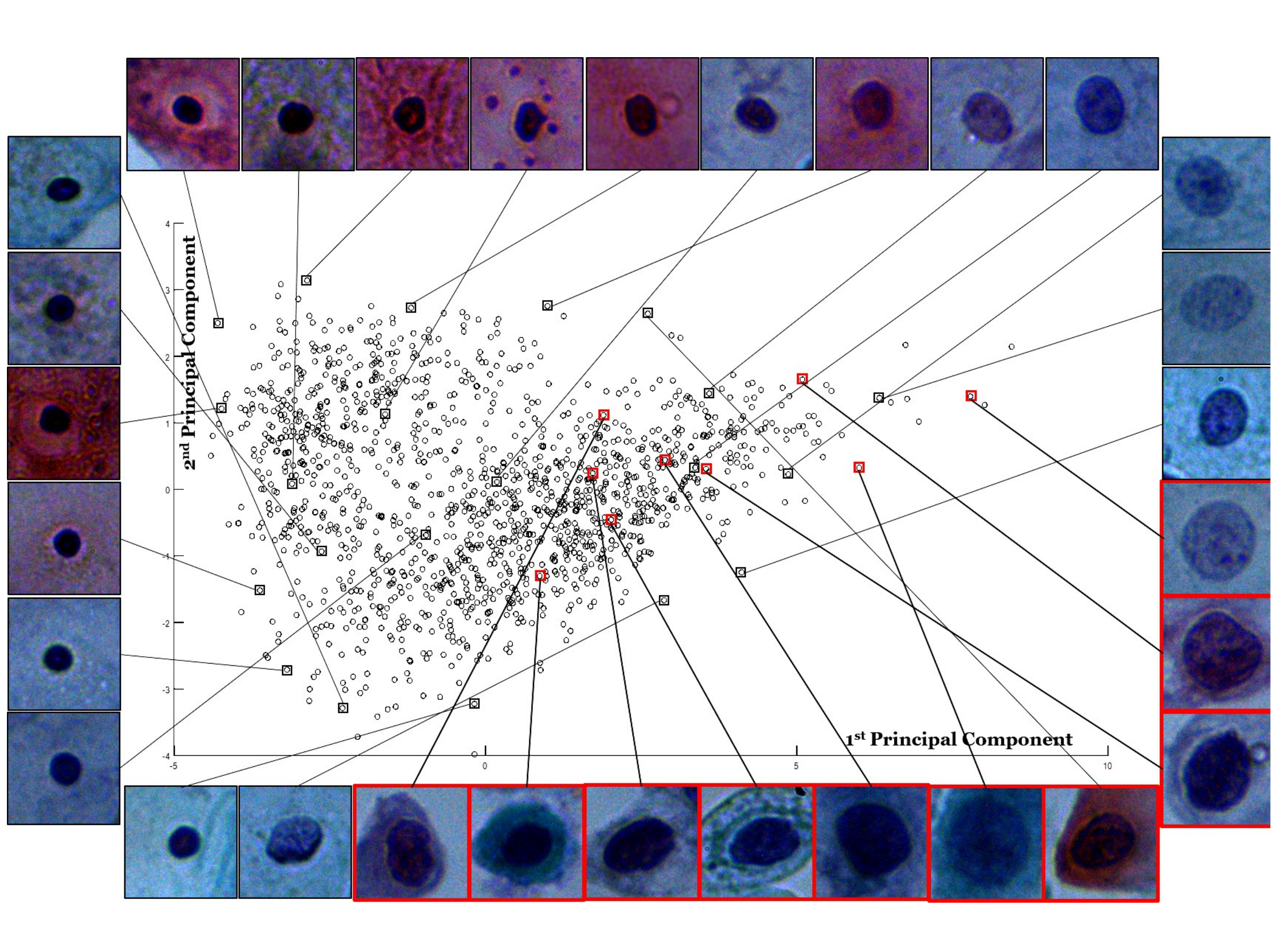}
\caption{Representation of cell data using first two PCA vectors obtained using data corresponding to $10$ measurements obtained from brightfield images of the $1450$ cell nuclei. The same representative brightfield images of $30$ cell nuclei as in Fig. \ref{fig:allpca} are also shown on the plot. The representation shows that the cells of similar type are not necessarily clustered together.}
\label{fig:bfpca}
\end{figure}

In order to understand the importance of phase information to this unsupervised organization of the cell image data, a second PCA analysis was performed with the numerical parameters calculated using the brightfield images alone. This is the information that is usually accessible to pathologists through typical microscopic imaging of the cell samples. The Kaiser-Meyer-Olkin score for the brightfield alone data for the $1450$ cells was found to be $0.64$ which is adequate but not as good as when phase parameters were included. The PCA plot corresponding to the brightfield parameters alone is shown in Fig. \ref{fig:bfpca}. For comparison we have shown the brightfield images of the same $30$ nuclei as in Fig. \ref{fig:allpca} on the plot and observe that they are not organized as per their type as in the PCA plot obtained with both brightfield as well as phase parameters. The cell types are mixed in the brightfield-only PCA plot. The abnormal cells (marked with red boundary) are for example not located in any one specific region of the plot. Further the PCA with brightfield parameters alone shows two large scale clusters with no apparent clinical value. 

\begin{figure}[tb]
\centering
\includegraphics[width = 0.9\textwidth]{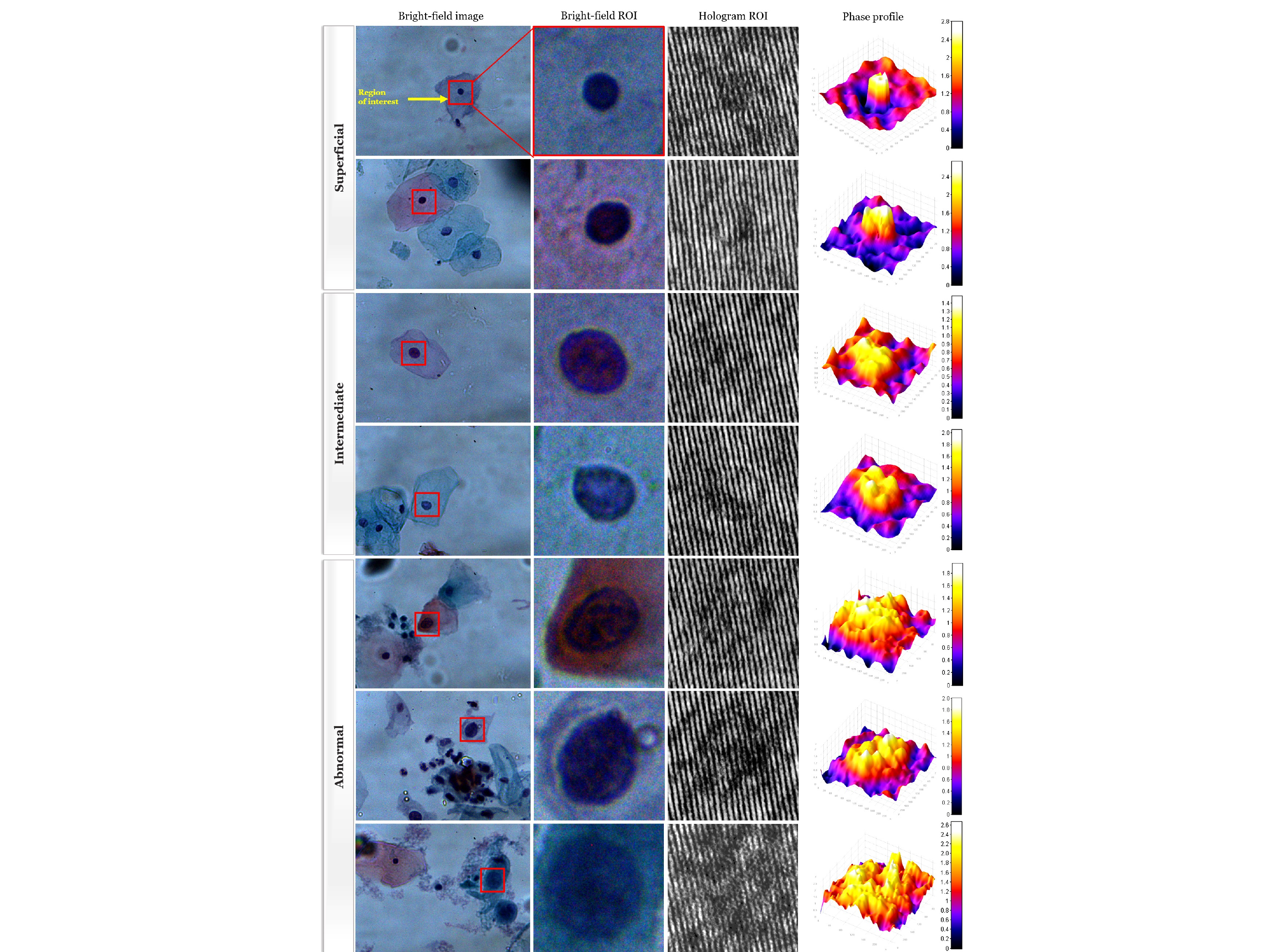}
\caption{Representative brightfield images depicting the various types of cervical cells and their morphology. As depicted, a region-of-interest is selected in the brightfield image centered on cell nucleus is shown in the left column. The corresponding local holograms are shown in the middle column. The corresponding phase profile of the nuclei rendered as surface plots is shown in the right column. The representative cells are selected from different regions of the PCA plot and the labels are provided by a pathologist. The phase profiles of nuclei show significant morphological changes for different kinds of cells.}
\label{fig:catalogue}
\end{figure}

The PCA analysis therefore clearly suggests that the phase information obtained with DHM is important in organizing the cell data. While identifying cell types among the abnormal cells will require a much larger data set consisting of sufficient cells of individual type that can be incorporated in the PCA analysis, the present result suggests that the additional non-redundant phase information about the cell nuclei may already considerably simplify the workflow for a pathologist and therefore has important clinical value. The PCA results are very interesting but do not clearly specify the physical information that the phase images bring into the analysis. In order to provide a visual interpretation of the phase images, we show brightfield as well as phase images of a few representative cell nuclei taken from different regions of the PCA plot in Fig. \ref{fig:catalogue}. The brightfield images of ROIs centered on $7$ different cell nuclei are shown in this figure along with their local hologram and the phase profile rendered as a surface plot. The phase profiles of these nuclei depict that the different classes of cells have noticeably different morphological structure.  In particular the phase images representing the abnormal cells show considerable fluctuations in the phase profile. The corresponding fringes in ROI of the hologram are also show higher degree of fringe modulation as expected. The nucleus of cancerous cells is known to undergo morphological changes in terms of nuclear size, shape and texture (e.g. indentations, folds and undulations) \cite{Zink}. Detailed association of phase profile of the nuclei with cell types is not within the scope of this paper but can be an interesting problem for future study.

The information in optical height profile which is available from phase but in principle not derivable from the brightfield data is thus seen to influence the PCA algorithm in a way that allows for better organization of cell nuclei as seen in Fig. \ref{fig:allpca}. 
\section{Conclusion}
Classification of cervical cells for screening and diagnostic purposes is an important problem representing a global healthcare challenge. A large number of qualitative parameters used by pathologists makes the problem of cell classification subjective. Automated analysis of cell images using typical brightfield microscopy images of cells requires labeling of a training set of cell image data which in itself is subjective in nature. An unsupervised organization of cell data is therefore highly desirable. In our study we used brightfield as well as phase images of cervical cells for exploring the feasibility of such unsupervised handling of cervical cell data. The phase images of the cells were obtained using a novel single-shot high resolution digital holographic microscopy (DHM) system. This system employs the traditional off-axis holography setup but uses an optimization algorithm for phase recovery with full diffraction-limited resolution that is not possible with conventional Fourier filtering algorithm. This low cost system does not require any moving parts (such as phase shifting elements) and the optimization methodology is proven to not compromise on the resolution and noise performance of phase imaging. A data set of $1450$ cell nuclei obtained from $10$ pap-smear slides was used in this study. For feature selection, we measured $21$ different morphological parameters using the brightfield as well as phase images of the cell nuclei. A visual inspection of the phase profiles of the cell nuclei show a clear distinction in their optical height morphology for different cell types. The unsupervised PCA analysis of the data consequently shows that the additional phase information which is not presently used in the clinical practice has tremendous value in organization of the cell data as per cell types. While a much larger data set may be required for further classification of normal or abnormal cells into futher sub-classes, the results indicate that the quantitative phase information that has largely been ignored in diagnostic pathology applications can potentially become an important clinical parameter for classification of cervical as well as other cell types. \\

\section*{Acknowledgements}
This work was partially supported by following grants: Department of Science and Technology India (ID/MED/34/2016), Biotechnology Industry Research Assistance Council (BIRAC), Department of Biotechnology India (BT/SBIRI1457/33/16). We also thank Holmarc Opto-Mechatronics Pvt. Ltd. for design and assembly work on the DHM product prototype system used for this work.\\\\\\\\\\

\end{document}